\documentclass[onecolumn,a4paper]{IEEEtran}
\IEEEoverridecommandlockouts
\usepackage[left=1.26cm,right=1.26cm,top=1.83cm,bottom=4.35cm]{geometry}
\usepackage{mleftright}       % fix to wrong spacing of \left-,
\mleftright                   % \middle- \right-commands 

\usepackage{graphicx}         % provides \includegraphics{...} to
\usepackage{booktabs} 

\usepackage[utf8]{inputenc}
\usepackage{amsmath,amssymb,amsfonts}
\usepackage{mathtools}
\usepackage{amsthm}
\usepackage{algorithmic}
\usepackage{ragged2e}
\usepackage{textcomp}
\usepackage{tikz}
\usepackage{caption}
\usepackage{cuted}
\usepackage{pgfgantt}
\usepackage{pdflscape}
\usepackage{pst-plot}
\usepackage{comment} 
\usepackage{cases}
\usepackage{nccfoots}
\usepackage{lineno}
\usepackage{graphicx}
\usepackage{float}
\usepackage[caption=false,font=footnotesize]{subfig}
\usetikzlibrary{spy}
\usetikzlibrary{positioning,calc}
\usetikzlibrary{decorations.pathmorphing,calc,shapes,shapes.geometric,patterns}
\usetikzlibrary{shapes.multipart}
\usepackage{xfrac}
\usepackage{colortbl}
\usepackage{cancel} 
\usepackage{bbm}
\usetikzlibrary{arrows,positioning,calc,intersections}
\usetikzlibrary{datavisualization.formats.functions}
\def\BibTeX{{\rm B\kern-.05em{\sc i\kern-.025em b}\kern-.08em
    T\kern-.1667em\lower.7ex\hbox{E}\kern-.125emX}}
    
\usepackage{romannum}
\usepackage{pgfplots}
\usepgfplotslibrary{fillbetween}
\usetikzlibrary{arrows, decorations.markings}
\usetikzlibrary{arrows.meta}

\newtheorem{theorem}{Theorem}
\newtheorem*{theorem*}{Theorem}
\newtheorem{lemma}[theorem]{Lemma}

\newtheorem{definition}[theorem]{Definition}

\newtheorem{remark}[theorem]{Remark}

\DeclarePairedDelimiterX{\infdivx}[2]{(}{)}{%
  #1\;\delimsize\|\;#2%
}

\pgfplotsset{compat=1.18}
\begin{document}
\title{An Achievable Rate-Distortion Region of\\ Joint Identification and Sensing\\ for Multiple Access Channels} 

% %%% Single author, or several authors with same affiliation:
% \author{%
%   \IEEEauthorblockN{Stefan M.~Moser}
%   \IEEEauthorblockA{ETH Zürich\\
%                     ISI (D-ITET)\\
%                     CH-8092 Zürich, Switzerland\\
%                     Email: moser@isi.ee.ethz.ch}
% }

%%% Several authors with up to three affiliations:

\author{
\IEEEauthorblockN{Yaning Zhao\IEEEauthorrefmark{1}\IEEEauthorrefmark{2}, Wafa Labidi \IEEEauthorrefmark{1}\IEEEauthorrefmark{2}\IEEEauthorrefmark{4}\thanks{\IEEEauthorrefmark{4}BMBF Research Hub 6G-life, Germany}, Holger Boche\IEEEauthorrefmark{1}\thanks{\IEEEauthorrefmark{3}Cyber Security in the Age of Large-Scale Adversaries–
Exzellenzcluster, Ruhr-Universit\"at Bochum, Germany}\IEEEauthorrefmark{4}\thanks{\IEEEauthorrefmark{6}{\color{black}{ Munich Center for Quantum Science and Technology (MCQST) }}}\IEEEauthorrefmark{7}\thanks{\IEEEauthorrefmark{7} {\color{black}{Munich Quantum Valley (MQV)}} }, Eduard Jorswieck\IEEEauthorrefmark{2}\IEEEauthorrefmark{5} \thanks{\IEEEauthorrefmark{5}BMBF Research Hub 6G-RIC, Germany}, and Christian Deppe\IEEEauthorrefmark{2}\IEEEauthorrefmark{4}} 
\IEEEauthorblockA{\IEEEauthorrefmark{1}\\Technical University of Munich,
\IEEEauthorrefmark{2}Technical University of Braunschweig\\
Email: yaning.zhao@tu-bs.de, wafa.labidi@tum.de, boche@tum.de, e.jorswieck@tu-bs.de, christian.deppe@tu-bs.de}
}

\maketitle

%%%%%% 
%% Abstract: 
%% If your paper is eligible for the student paper award, please add
%% the comment "THIS PAPER IS ELIGIBLE FOR THE STUDENT PAPER
%% AWARD." as a first line in the abstract. 
%% For the final version of the accepted paper, please do not forget
%% to remove this comment!
%%

%====================== abstract =========================
\begin{abstract}
In contrast to Shannon transmission codes, the size of identification (ID) codes for discrete memoryless channels (DMCs) experiences doubly exponential growth with the block length when randomized encoding is used. Additional enhancements within the ID paradigm can be realized through supplementary resources such as quantum entanglement, common randomness (CR), and feedback. Joint transmission and sensing demonstrate significant benefits over separation-based methods. Inspired by the significant impact of feedback on the ID capacity, our work delves into the realm of joint ID and sensing (JIDAS) for state-dependent multiple access channels (SD-MACs) with noiseless strictly casual feedback. Here, the senders aim to convey ID messages to the receiver while simultaneously sensing the channel states. We establish a lower bound on the capacity-distortion region of the SD-MACs. An example shows that JIDAS outperforms the separation-based approach.
\end{abstract}
%\textit{A detailed version with all proofs, explanations, and more discussions can be found in {\color{red} [perhapsanarxivversion]}}.

%==================== introduction ========================
\section{Introduction}
\label{sec:introduction}
%%%%%% cite Shannon %%%%%%
Shannon's groundbreaking work \cite{shannon1948mathematical} laid the foundation for the transmission problem, revealing that the maximum number of reliably transmitted messages grows exponentially with block length.
%%%%%% cite application / cite Jaja %%%%%%
 In today's technological landscape including machine-to-machine systems \cite{boche2018secure}, digital watermarking \cite{moulin2001role,ahlswede2006watermarking,steinberg2001identification}, industry 4.0 \cite{lu2017industry}, and 6G communication systems \cite{fettweis20226g,cabrera20216g}, the demand for improving data rates, latency, and security \cite{fettweis20226g} is increasing. Many applications utilize Shannon's transmission scheme, requiring receivers to decode all messages. However, this approach proves inefficient in various scenarios. Therefore, Ahlswede and Dueck \cite{ahlswede1989identification}, building on Jaja's work \cite{ja1985identification}, introduced the Identification (ID) scheme. Here, the receiver aims to decide whether a special message has been transmitted or not.
 Of course, the sender does not know which message is actually interesting to the receiver.

 A groundbreaking result is demonstrated: when a randomized encoding scheme is employed, the size of identities experiences a doubly exponential growth with the block length. With only a deterministic coding scheme, despite scaling similarly to the transmission code, the ID rate still far surpasses the transmission rate for DMCs.
%%%%%% cite other ID works %%%%%%%
Research demonstrates seamless integration of information-theoretic security into identification without added secrecy costs \cite{ahlswede1995new,labidi2020secure}. Advancements in ID schemes can utilize resources such as quantum entanglement \cite{boche2019secure,pereg2022identification}, common randomness (CR) \cite{ezzine2024common,ezzine2021common,labidi2022common}, and feedback \cite{ahlswede1989identificationfeedback}. Quantum entanglement is often more effective than CR.
%%%%%% cite Alhswede IDF %%%%%%
The study in \cite{wolfowitz2012coding} demonstrates that feedback does not increase transmission capacity for DMCs. However, feedback can increase ID capacity in noisy channels \cite{ahlswede1989identificationfeedback}. 
Identification in particular and post Shannon communication in general is a key approach to effectively implement important future applications to meet the cost and energy reduction challenges for future network operators \cite{Schwenteck2023}.

%%%%%%%%%%%%%%%%%%%%%%%%%%% JID %%%%%%%%%%%%%%%%%%%%%%%%%%%%%%%%%%%%%%%
\indent In upcoming 6G networks, precise sensing of state information (SI) is vital for optimal performance \cite{bourdoux20206g, Fettweis2021}. 
Traditionally, message transmission and sensing are handled separately, with resources divided for either estimating SI or communication through time-sharing. However, there is a trend towards merging them into a unified system and platform \cite{zheng2019radar,liu2020joint}. Numerous recent studies have addressed the challenge of joint transmission and sensing \cite{sturm2011waveform,bliss2014cooperative,bica2015opportunistic,huang2015radar}. For example, the single-user DMCs have been explored in \cite{kobayashi2018joint}. In their study, messages are transmitted, while an estimator on the sender's side senses SI via causal feedback. They introduce a capacity-distortion trade-off, where the capacity is the supremum of achievable message transmission rates when ensuring that the distortion of sensing remains below a threshold value, representing the maximum distortion tolerable. Moreover, the work by \cite{9834554} extended this model to MACs, and \cite{10153971} addressed it within the context of broadcast channels (BC). In joint identification and sensing (JIDAS), using a noiseless feedback link benefits both the estimator and encoder. In \cite{labidi2023joint}, they study JIDAS with noiseless feedback over single-user DMCs, finding a lower bound of the deterministic ID capacity-distortion trade-off. 
The combination of Post Shannon communication and sensing can be expected to achieve additional synergies in the area of cost reduction and energy consumption reduction for future grid operators and users \cite{Schwenteck2023}. JDAS is the first example of this combination. Central research questions are still completely open \cite{Fettweis2021}.
\\
%%%%%% MAC %%%%%%
\indent The MAC models multiple senders communicating with a single receiver. A typical MAC scenario is the up-link of cellular systems, where multiple devices send data to a central base station. Initial characterization of the message transmission problem via a MAC was done by \cite{liao1972multiple}, with further details provided in \cite{el2011network}. The capacity of deterministic ID for MACs is explored in \cite{rosenberger2023deterministic}, while \cite{ahlswede2008general} investigates randomized ID capacity. Quantum entanglement's enhancement of randomized ID via a MAC is shown in \cite{diadamo2019simultaneous}, with further advantages of perfect feedback discussed in \cite{ahlswede1971multi}.
To the best of our knowledge, the JIDAS problem for SD-MACs has not been addressed in the existing literature. In this work, we establish a lower bound on the ID capacity-distortion region for a two-sender state-dependent multiple access channel (K-SD-MAC) and demonstrate that our joint approach outperforms the separation-based scenario.\\
\indent \textit{Outline}: The remainder of the paper is structured as follows. In Section \ref{sec: models&results}, we introduce our system model and present the main results. In Section \ref{sec: proof}, we establish proof of the lower bound on the ID capacity-distortion region for a K-SD-MAC. In Section \ref{sec: examples}, we provide an example to show the benefits of our joint approach. Section \ref{sec: conclusion} concludes the paper.

%================ system model & main reults ==============
\section{System Model and Main Results}
\label{sec: models&results}
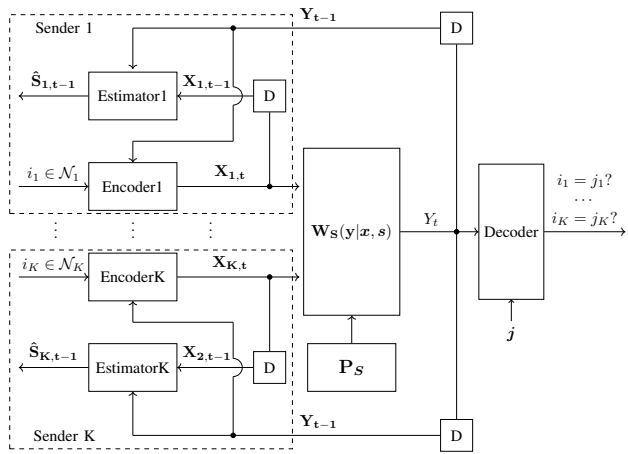
\begin{figure}
    \centering
    \hspace*{-2.15em}
    \scalebox{0.6}{
    \tikzstyle{block} = [draw, fill=white, rectangle, minimum height=3em, minimum width=5.5em]
    \tikzstyle{lblock}=[draw, fill=white, rectangle, minimum height=8.5em, minimum width=4em]
    \tikzstyle{channel}=[draw, fill=white, rectangle, minimum height=10.5em, minimum width=6em]
    \tikzstyle{point} = [draw, fill, circle, inner sep=1pt, outer sep=0pt]
    \tikzstyle{input} = [coordinate]
    \tikzstyle{output} = [coordinate]
    \tikzstyle{pinstyle} = [pin edge={to-,thin,black}]
    \tikzstyle{box}=[draw, rectangle,minimum height=12.5em, minimum width=6.2cm]
    \begin{tikzpicture}[auto, node distance=3cm]
    \tikzstyle{smallblock}=[draw,fill=white, rectangle, minimum height =2em, minimum width=2em]
        \node[input](source1) at (0,0){};
        \node[block,right of=source1,node distance=2.5cm](encoder1){Encoder1};
        \node[block,above of=encoder1,node distance=2cm](estimator1){Estimator1};
        \node[input,left of=estimator1, node distance=2.5cm](estoutput1){};
        \node[input,below of=source1,node distance=2cm](source2){};
        \node[block,below of=encoder1,node distance=2cm](encoder2){EncoderK};
        \node[block,below of=encoder2,node distance=2cm](estimator2){EstimatorK};
        \node[input,left of=estimator2, node distance=2.5cm](estoutput2){};

        \node[input,below of=encoder1,node distance=1cm](input){};
        \node[channel,right of=input,node distance=4.8cm,align=center](channel){$\mathbf{W_S(y|\boldsymbol{x},\boldsymbol{s})}$};
        \node[input,right of=encoder1,node distance=3.65cm](channel_input_1){};
        \node[point,right of=encoder1,node distance=3cm](est1){};
        \node[smallblock,above of=est1, node distance = 2cm](est1_1){D};
        \node[input,right of=encoder2,node distance=3.65cm](channel_input_2){};
        \node[point,right of=encoder2,node distance=3cm](est2){};
        \node[smallblock,below of=est2, node distance = 2cm](est2_1){D};
        \node[lblock,right of=channel,node distance=3.5cm](decoder){Decoder};
        \node[input,below of=decoder,node distance=1.5cm,pin={[pinstyle]below:$\boldsymbol{j}$}](i'j'){};
        \node[output,right of=decoder,node distance =2.5cm](sink){};
        \node[block,below of=channel,node distance=3cm](state){\large$\mathbf{P_{\boldsymbol{S}}}$};

        \node[point,right of=channel,node distance=2.3cm](f1){};
        \node[smallblock,above of=f1,node distance=4.5cm](f2){D};
        \node[input,above of=estimator1,node distance=1.5cm](f3){};
        \node[input,above of=estimator1,node distance=0.6cm](f4){};
        \node[point,right of=f3,node distance=2.2cm](f8){};
        \node[input,below of=f8, node distance =1.3cm](f9){};
        \node[input,below of=f9, node distance =0.4cm](f10){};
        \node[input,below of=f10, node distance =0.8cm](f11){};
        \node[input,above of =encoder1, node distance=1cm](f12){};

        \node[smallblock,below of=f1,node distance=4.5cm](f5){D};
        \node[input,below of=estimator2,node distance=1.5cm](f6){};
        \node[point,right of=f6,node distance=2.2cm](f13){};
        \node[input,above of=f13, node distance =1.3cm](f14){};
        \node[input,above of=f14, node distance =0.4cm](f15){};
        \node[input,above of=f15, node distance =0.8cm](f16){};
        \node[input,below of =encoder2, node distance=1cm](f17){};
        \node[] at (0.8,-1) {\rotatebox{-90}{\textbf{$\cdots$}}};
        \node[] at (2.5,-1) {\rotatebox{-90}{\textbf{$\cdots$}}};
        \node[] at (4.2,-1) {\rotatebox{-90}{\textbf{$\cdots$}}};
        
        \node[input,below of=encoder2,node distance=0.6cm](f7){}; 
        \draw [->] (source1) -- node[name=i] {$i_1\in\mathcal{N}_1$} (encoder1);
        \draw[->](source2) --  node[name=j] {$i_K\in\mathcal{N}_K$} (encoder2);
        \draw[->](encoder1) -- node[name=x1,align=right] {$\mathbf{X_{1,t}}\quad$}(channel_input_1);
        \draw[->](encoder2) -- node[name=x2,align=right] {$\mathbf{X_{K,t}}\quad$}(channel_input_2);
        \draw[->](channel) -- node[name=y]{$Y_t\quad$}(decoder);
        \draw[->](decoder) -- node[name=result,align=center]{$i_1=j_1?$\\$\cdots$\\$i_K=j_K?$}(sink);
        \draw[->](state)--(channel);
        \draw[-](f1)--(f2);
        \draw[-](f2)--node[above]{$\quad \quad \quad \quad \mathbf{Y_{t-1}}$}(f3);
        \draw[->](f3)--(f4);
        \draw[-](f1)--(f5);
        \draw[-](f5)--node[above]{$\quad \quad \quad \quad \mathbf{Y_{t-1}}$}(f6);
        \draw[->](f6)--(estimator2);
        \draw[-](est1)--(est1_1);
        \draw[-](est2)--(est2_1);
        \draw[->](est1_1)--node[above]{$\mathbf{X_{1,t-1}\quad}$}(estimator1);
        \draw[->](est2_1)--node[above]{$\mathbf{X_{2,t-1}\quad}$}(estimator2);
        \draw[->](estimator1)--node[above]{$\mathbf{\hat{S}_{1,t-1}}$}(estoutput1);
        \draw[->](estimator2)--node[above]{$\mathbf{\hat{S}_{K,t-1}}$}(estoutput2);
        \draw[-](f8)--(f9);
        \draw[black](f9) arc(90:-90:0.2cm);
        \draw[-](f10)--(f11);
        \draw[-](f11)--(f12);
        \draw[->](f12)--(encoder1);
        \draw[-](f13)--(f14);
        \draw[black](f14) arc(-90:90:0.2cm);
        \draw[-](f15)--(f16);
        \draw[-](f16)--(f17);
        \draw[->](f17)--(encoder2);
        \node[box, dashed] at (2.9,1.6) (sender) {};
        \node[box, dashed] at (2.9,-3.6) (sender) {};
        \node at (1,3.5) {Sender 1};
        \node at (1,-5.5) {Sender K};

    \end{tikzpicture}}
    \caption{K-SD-MAC with noiseless causal feedback}
    \label{fig:system_model}
    \vspace{-1.8em}
\end{figure}

Consider a K-Sender-SD-MAC (K-SD-MAC) 
$(\boldsymbol{\mathcal{X}}\times\boldsymbol{\mathcal{S}}, W_S(y|\boldsymbol{x},\boldsymbol{s}),\mathcal{Y})$ with noiseless causal feedback as depicted in Fig. \ref{fig:system_model}. The K-SD-MAC is described by a conditional probability distribution $W_{S}(Y|\boldsymbol{X},\boldsymbol{S})$. The channel input tuple $\boldsymbol{X}=\left(X_{1},\cdots,X_{K}\right)$, the channel state tuple $\boldsymbol{S}=\left(S_{1},\cdots,S_{K}\right)$, and the channel output $Y$ take values in finite sets $\boldsymbol{\mathcal{X}}=\mathcal{X}_{1}\times\cdots\times\mathcal{X}_{K}$, $\boldsymbol{\mathcal{S}}=\mathcal{S}_{1}\times\cdots\times\mathcal{S}_{K}$, and $\mathcal{Y}$, respectively.
Given blocklength $n$, the state tuples $\boldsymbol{S}_t$ for all $t=1,\cdots,n$ are i.i.d. according to the joint distribution $P_{\boldsymbol{S}}$. We assume that the inputs $\boldsymbol{X}_{t}$, and the states $\boldsymbol{S}_t$ are statistically independent for all $t=1,\cdots,n$. The channel transition probability is memoryless, i.e., $W_S(y^n|\boldsymbol{x}^n,\boldsymbol{s}^n)=\prod_{t=1}^n{W(y_t|\boldsymbol{x}_{t},\boldsymbol{s}_{t})}$. The notation $D$ represents a one-symbol-time delay. The feedback is noiseless and strictly causal, that is, at time $t$ only $Y^{t-1}=[Y_1,\cdots,Y_{t-1}]$ is available at encoders and estimators.
We consider the following average channel:
\begin{align*}
    W_{S\_avg}(y|\boldsymbol{x})=\sum_{\boldsymbol{s}\in\boldsymbol{\mathcal{S}}}P_{\boldsymbol{S}}(\boldsymbol{s})W_S(y|\boldsymbol{x},\boldsymbol{s}), 
\end{align*}
for all $\boldsymbol{x}\in\boldsymbol{\mathcal{X}}$, and $y\in\mathcal{Y}$.

Each encoder $k\in\mathcal{K}$ has a set of identities $\mathcal{N}_{k}=\left\{1,\cdots,N_k\right\}$. They independently send identities $i_k\in\mathcal{N}_k$ to the receiver. Each identity $i_k$ is encoded according to a strictly-causal noiseless feedback sequence $y^{t-1}$. The feedback encoding functions $\boldsymbol{f}_{k,i_{k}}$ is defined as follows.
\begin{definition}
    A feedback encoding function for identity $i_k\in\mathcal{N}_k$ is a vector-valued function
    \begin{align*}
        \boldsymbol{f}_{k,i_k}=\left[f_{k,i_k}^1,\cdots,f_{k,i_k}^n\right],
    \end{align*}
    where $f_{k,i_k}\in\mathcal{X}_{k}$, and for $t\in\left\{2,\cdots,n\right\}$, $f_{k,i_k}^t:\mathcal{Y}^{t-1}\mapsto \mathcal{X}_k$.\\
    
    We denote the set of $\boldsymbol{f}_{k,i_k}$ with length $n$ as $\mathcal{F}_k^n$.
\end{definition}
In the following, we define deterministic and randomized IDF codes for a K-SD-MAC.
\begin{definition}
    An $(n,\boldsymbol{N},\lambda)$ deterministic IDF code with $\lambda\in(0,\frac{1}{2})$ for K-SD-MAC $W_S$ is a system $\left\{(\boldsymbol{f}_{\boldsymbol{i}},\mathcal{D}_{\boldsymbol{i}})|\boldsymbol{i}\in\boldsymbol{\mathcal{N}}\right\}$, where
    \begin{align*}
       \boldsymbol{f}_{\boldsymbol{i}}\in\boldsymbol{\mathcal{F}}^n,\quad \mathcal{D}_{\boldsymbol{i}}\subset \mathcal{Y}^n.
    \end{align*}
    The probabilities of type I error and type II error for identity tuple $\boldsymbol{\mathcal{N}}$ satisfy
\begin{align*}
        P_{e,1}\left(\boldsymbol{i}\right)&\triangleq W_{S\_avg}\left(\mathcal{D}^c_{\boldsymbol{i}}|\boldsymbol{f}_{\boldsymbol{i}}\right)\le \lambda,\quad \forall \boldsymbol{i}\in\boldsymbol{\mathcal{N}},\\
        P_{e,2}\left(\boldsymbol{i},\tilde{\boldsymbol{i}}\right)&\triangleq W_{S\_avg}(\mathcal{D}_{\boldsymbol{\tilde{i}}}|\boldsymbol{f}_{\boldsymbol{i}})\le \lambda,\quad \forall \boldsymbol{i},\tilde{\boldsymbol{i}}\in\boldsymbol{\mathcal{N}},\tilde{\boldsymbol{i}}\ne \boldsymbol{i}.
    \end{align*}
\end{definition}

\begin{definition}
    $\quad$ An $(n,\boldsymbol{N},\lambda)$ randomized IDF code with $\lambda\in(0,\frac{1}{2})$ for K-SD-MAC $W_S$ is a system $\left\{(Q(\cdot|\boldsymbol{i}),\mathcal{D}_{\boldsymbol{i}})|\boldsymbol{i}\in\boldsymbol{\mathcal{N}}\right\}$ with
    \begin{align*}
        Q(\cdot|\boldsymbol{i})\in \mathcal{P}(\boldsymbol{\mathcal{F}}^n),\quad \mathcal{D}_{\boldsymbol{i}}\subset \mathcal{Y}^n,
    \end{align*}
    where $Q(\cdot|\boldsymbol{i})=\prod_{k=1}^{K}Q_k(\cdot|i_k)$, and $Q_k(\cdot|i_k)\in\mathcal{P}(\mathcal{F}_k^n)$.
    
    The probabilities of type I error and type II error for identity tuple $\boldsymbol{\mathcal{N}}$ satisfy
    \begin{align*}
        P_{e,1}\left(\boldsymbol{i}\right)&\triangleq\sum_{\boldsymbol{f}\in \boldsymbol{\mathcal{F}}^n}Q(\boldsymbol{f}|i)W_{S\_avg}^n(\mathcal{D}_{\boldsymbol{i}}^c|\boldsymbol{f})\le\lambda,\quad
        \forall \boldsymbol{i}\in\boldsymbol{\mathcal{N}},\\ 
        P_{e,2}\left(\boldsymbol{i},\tilde{\boldsymbol{i}}\right)&\triangleq\sum_{\boldsymbol{f}\in \boldsymbol{\mathcal{F}}^n}Q(\boldsymbol{f}|i)W_{S\_avg}^n(\mathcal{D}_{\boldsymbol{\tilde{i}}}^c|\boldsymbol{f})\le \lambda,\\
        \forall \boldsymbol{i},\boldsymbol{\tilde{i}}\in\boldsymbol{\mathcal{N}}&, \boldsymbol{i}\neq\boldsymbol{\tilde{i}}.
\end{align*}
\end{definition}
Simultaneously, estimator $k$ senses its respective states $S_k$, generating estimation $\hat{S}_{k,t-1}$ with respect to feedback symbol $Y_{t-1}$ and input symbol $X_{k,t-1}$. The performance of estimators is evaluated based on the expected distortion over $(S_k,\hat{S_k})$, i.e.,
\begin{align}
    \mathbb{E}\left[d(S_k^n,\hat{S}_k^n)\right]=\frac{1}{n}\sum_{t=1}^{n}{\mathbb{E}\left[d(S_{k,t},\hat{S}_{k,t})\right]},
    \label{eq:expected distortion}
\end{align}
where $d(\cdot,\cdot):\hat{\mathcal{S}}_k\times\mathcal{S}_k\mapsto \mathbb{R}$ is a general distortion function, e.g., Hamming distance, or mean square error (MSE).

We define the estimation functions $h_k:\mathcal{X}_k\times\mathcal{Y}\mapsto\mathcal{S}_k$, and denote the sets of all $h_k$ as $\mathcal{H}_k$. Without loss of generality, deterministic estimators $h^*_k(\cdot,\cdot)$ can be employed, defined as follows:
\begin{align*}
    h^*_k(x_k,y)=\underset{{h_k\in\mathcal{H}_k}}{\arg\min} \mathbb{E}_{S_k}\left[d(S_k,h(X_k,Y))|X_k=x_k,Y=y\right].
\end{align*}

 We define the minimal distortion for each input symbol and input distribution, respectively, achieved by the deterministic estimator described above.
\begin{definition}
    For each input-symbol $x_k\in\mathcal{X}_k$, the minimal distortion $d_k^*(x_k)$ is given by
    \begin{align*}
        d_k^*(x_k)=\mathbb{E}_{S_kY}\left[d(S_k,h_k^*(X_k,Y))|X_k=x_k\right].
    \end{align*}
    Similarly, for each input distribution $P_k\in\mathcal{P}(\mathcal{X}_k)$, the minimal distortion $d_k^*(P_k)$ is given by
    \begin{align*}
        d_k^*(P_k)=\sum_{x_k\in\mathcal{X}_k}P_k(x_k)d_k^*(x_k).
    \end{align*}
\end{definition}
\noindent Thus, for the deterministic coding strategies $\boldsymbol{f}_i$ and $\boldsymbol{g}_j$, we can express the expected distortions \eqref{eq:expected distortion} as follows:
\begin{align*}
    \bar{d}^n_k
    &=\frac{1}{N_k}\sum_{i_k=1}^{N_k}\frac{1}{n}\sum_{t=1}^n d_k^*(f_{k,i_k}^t).
\end{align*}
Similarly, for randomized coding strategy $Q(\cdot|\boldsymbol{i})$, we can represent the average distortion as follows:
\begin{align*}
    \bar{d}_k^n=\frac{1}{N_k}\sum_{i_k=1}^{N_k}\sum_{\boldsymbol{f}_k\in\mathcal{F}_k^n}Q_k(\boldsymbol{f}_k|i_{k})\frac{1}{n}\sum_{t=1}^n d_k^*\left(f_{k,i_k}^t\right).
\end{align*}
\begin{definition}
    \begin{enumerate}
        \item A rate-distortion tuple $\boldsymbol{R}(\boldsymbol{D})\triangleq(R_1(\boldsymbol{D}),\cdots,R_K(\boldsymbol{D}))$ with vector-valued constant $\boldsymbol{D}=\left(D_1,\cdots,D_K\right)$ representing the maximum tolerated distortions is said to be achievable, if there exists an $(n,\boldsymbol{\mathcal{N}},\lambda)$ IDF code for K-SD-MAC $W_S$, such that for all $k\in\mathcal{K}$, $\limsup_{n\to\infty}\bar{d}^n_k\le D_k$.
        \item The ID capacity-distortion region $\mathcal{C}(\boldsymbol{D})=\left\{R_1(\boldsymbol{D}),\cdots,R_K(\boldsymbol{D})\right\}$ is defined as the closure of all achievable rate-distortion tuples.
    \end{enumerate}
\end{definition}
\begin{theorem}
If all senders can transmit with positive rates, then the deterministic ID capacity-distortion region $\mathcal{C}_{ID}^d(\boldsymbol{D})$ of a K-SD-MAC $W_S$ is lower-bounded by
\begin{align}
\label{eq.JdIDAS}
    \begin{split}
        \mathcal{C}_{ID}^{d}(\boldsymbol{D})=
    \Biggl\{\boldsymbol{R}(\boldsymbol{D}):R_k(\boldsymbol{D})\geq
    &\max_{\boldsymbol{x}\in\boldsymbol{\mathcal{X}}^{D}}H(W_{S\_avg}(\cdot|\boldsymbol{x}))\Biggr\},
    \end{split}
\end{align}
where $\boldsymbol{\mathcal{X}}^{D}=\mathcal{X}_{1}^{D_{1}}\times \cdots \times\mathcal{X}_{K}^{D_{K}}$, and $\mathcal{X}_{k}^{D_k}=\left\{x_k\in\mathcal{X}_k|d_k^*(x_k)\le D_k\right\}$ representing the sets of admissible input symbols.
\label{thm.JdIDAS}
\end{theorem}

\begin{theorem}
    If all senders can transmit with positive rates, then the randomized ID capacity-distortion region $\mathcal{C}_{ID}^r(\boldsymbol{D})$ of a K-SD-MAC $W_S$ is lower-bounded by
    \begin{align}
    \label{eq.JrIDAS}
    \begin{split}
        \mathcal{C}_{ID}^r(\boldsymbol{D})=
    \Biggl\{\boldsymbol{R}(\boldsymbol{D}):R_k(\boldsymbol{D})\geq
    &\max_{\boldsymbol{P}\in\boldsymbol{\mathcal{P}}^{\boldsymbol{D}}}H\left(\boldsymbol{P}W_{S\_avg}\right)\Biggr\},
    \end{split}
    \end{align}
    where $\boldsymbol{\mathcal{P}}^{\boldsymbol{D}}=\mathcal{P}_{1}^{D_1}\times\cdots\times\mathcal{P}_K^{D_K}$, and $\mathcal{P}_{k}^{D_k}=\left\{P_k\in\mathcal{P}(\mathcal{X}_k)|d_k^*(P_k)\le D_k\right\}$ denoting the set of admissible input distributions. $W_{S\_avg}(y)=\sum_{\boldsymbol{x}\in\boldsymbol{\mathcal{X}}}\boldsymbol{P}(\boldsymbol{x})W_{S\_avg}(y|\boldsymbol{x})$, for all $y\in\mathcal{Y}$ is a distribution over $\mathcal{Y}$.
\label{thm.JrIDAS}
\end{theorem}
\begin{remark}
As long as all senders can transmit with positive rates, there is no trade-off between two senders. Moreover, the weaker sender (with lower transmission capacity) can achieve the same rate as the stronger sender.
\end{remark}
\begin{remark}
As long as all senders can transmit with positive rates, there is no trade-off between two senders. Moreover, the weaker sender (with lower transmission capacity) can achieve the same rate as the stronger sender.
\end{remark}

%========================== Proof ========================
\section{Proof}
\label{sec: proof}
In this section, we provide the proofs of Theorem \ref{thm.JdIDAS} and Theorem \ref{thm.JrIDAS}. It suffices to show that the above-mentioned rate-distortion regions are achievable for the average channel $W_{S\_avg}$.
\subsection{Proof of Theorem \ref{thm.JdIDAS}}
We extend the encoding scheme for IDF through single-user channels, as outlined in \cite{ahlswede1989identificationfeedback}, to our JIDAS problem via a K-SD-MAC. Consider an $(m,\boldsymbol{\mathcal{N}},\lambda)$ deterministic IDF code, where $m=n+\lceil\sqrt{n}\rceil$. The encoders use first $n$ bits to generate common randomness by sending $\boldsymbol{x}^{*n}$, where
\begin{align}
    \boldsymbol{x}^{*}=\underset{\boldsymbol{x}\in \boldsymbol{\mathcal{X}}^{\boldsymbol{D}}}{\arg\max} H(W_{S\_avg}(\cdot|\boldsymbol{x})).
    \label{eq:x1*x2*}
\end{align}
The sequences $y^n$ sharing among encoders, estimators (via noiseless feedback links) and decoder (via channel) can be regarded as outcomes of random experiments $\left(\mathcal{Y}^n,W_{S\_avg}^n(Y^n|\boldsymbol{x}^{*n})\right)$. However, it is not uniformly distributed across the sample space, presenting challenges in constructing codes. The theorem about typical sequences provides solutions to address this concern. We denote $\mathcal{D}^*=\mathcal{T}_{\epsilon}^n(W_{S\_avg}(\cdot|\boldsymbol{x}^{*}))$ as the set of typical sequences and $e=\mathcal{Y}/\mathcal{D}^*$ as the set of errors. By choosing small $\epsilon$ and sufficient large $n$, we have the following lemma:
\begin{lemma}
\label{lemma:typ}
    Suppose $y^n\in \mathcal{D}^*$, and $Y^n$ is emitted by $W_{S\_avg}(\cdot|\boldsymbol{x}^{*})$, then
    \begin{align*}
        W_{S\_avg}^n(y^n|\boldsymbol{x}^{*n})&\doteq 2^{-nH(W(\cdot|\boldsymbol{x}^{*}))},\\
        |\mathcal{D}^*|&\doteq 2^{nH(W(\cdot|\boldsymbol{x}^{*}))},\\
        W_{S\_avg}^n(\mathcal{D}^{*}|\boldsymbol{x}^{*n})& \doteq1,
    \end{align*} 
    where $\doteq$ denotes the asymptotic equivalence in $n$.
\end{lemma}
\noindent Thus, without loss generality, we convert the random experiment to a uniform one $\left(\mathcal{D}^*,W_{S\_avg}(y|\boldsymbol{x})\right)$. 

Next, we introduce families of "coloring functions", denoted as $\left\{F_{k,i_k}|i_k=1,\cdots,N_k\right\}$ for all $k\in\mathcal{K}$. Each function $F_{k,i_k}:\quad \mathcal{D}^* \mapsto \mathcal{M}_{k}=\left\{1,\cdots,M_k\right\}$ maps the sequence $y^n\in\mathcal{D}^*$ to a "color" $l_k\in\mathcal{M}_k$. We denote the tuple of "coloring functions" as $\boldsymbol{F}_{\boldsymbol{i}}=\left(F_{1,i_1},\cdots,F_{K,i_K}\right)$, the tuple of "colors" as $\boldsymbol{l}=\left(l_1,\cdots,l_K\right)$, the tuple of the color sets as $\boldsymbol{\mathcal{M}}=\left(\mathcal{M}_1,\cdots,\mathcal{M}_K\right)$, and their cardinality as $\boldsymbol{M}=\left(M_1,\cdots,M_K\right)$. We uniformly and randomly generate the functions, i.e., $Pr\left[F_{k,i_k}(y^n)=l_k\right]=\frac{1}{M_k}$. The mappings $F_{k,i_k}$ serve as prior knowledge for all encoders and the decoder.

After receiving $y^n\in\mathcal{D}^*$ via feedback, the encoders calculate and transmit $F_{k,i_k}(y^n)$, using a "normal" $\left(\lceil\sqrt{n}\rceil,\boldsymbol{M},2^{-\lceil\sqrt{n}\rceil\delta}\right)$ transmission code denoted as $\mathcal{C}'=\left\{\left(\boldsymbol{c}(\boldsymbol{l}),\mathcal{D}'_{\boldsymbol{l}}\right)\big|\boldsymbol{l}\in\mathcal{M}\right\}$. In the following process of analysis, we use $\boldsymbol{c}=\left(\boldsymbol{c}_1,\cdots,\boldsymbol{c}_K\right)$ to denote $\boldsymbol{c}\left(\boldsymbol{F}_{\boldsymbol{i}}(y^n)\right)$. Notably, we need to choose the code words under distortion constraints, i.e., $\bar{d}_k(\boldsymbol{c})=\frac{1}{\lceil\sqrt{n}\rceil}\sum_{t=1}^{\lceil\sqrt{n}\rceil}d_k^*(c_{k,t})\le D_k$ for all $k\in\mathcal{K}$. If $y^n\in e$, an error is declared. But according to Lemma \ref{lemma:typ}, this error occurs with a probability close to $0$. The decoder first computes $F_{i'_k}(y^n)$. Then, it decodes $\hat{{F}_{i_k}}$ based on $y_{n+1}^m$. If ${F}_{i'_k}(y^n)=\hat{F}_{i_k}$, it concludes $i_k=i'_k$; otherwise, $i_k\neq i'_k$.

In conclusion, for feedback sequences $y^n\in\mathcal{D}^*$, our deterministic IDF code is defined as
\begin{align*}
    \boldsymbol{f}_{k,i_k} &= \left[x_k^{*n}, \boldsymbol{c}_k\right],\quad x_k^{*}\in\mathcal{X}_k^{D_k},\quad\bar{d}_k(\boldsymbol{c}_k)\le D_k,\quad \forall k\in\mathcal{K},\\
    \mathcal{D}_{\boldsymbol{i}}&=\bigcup_{y^n\in \mathcal{D}^*}y^n\times \mathcal{D}'_{\boldsymbol{F}_{\boldsymbol{i}}(y^n)},\quad \forall \boldsymbol{i}\in\boldsymbol{\mathcal{N}},
\end{align*}
where $\boldsymbol{x}_k$ follows \eqref{eq:x1*x2*}.

For all $\boldsymbol{i}\in\boldsymbol{\mathcal{N}}$, the type I error probability at Sender k $P_{e,1}(\boldsymbol{i})$ can be upper-bounded by:
\begin{align*}
    &W_{S\_avg}^{m}(\mathcal{D}_{\boldsymbol{i}}^c|\boldsymbol{f}_{\boldsymbol{i}})\\
     &=\sum_{y^n\in \mathcal{D}^*}W_{S\_avg}^n\left(y^n|\boldsymbol{x}^{*n}\right)\cdot W_{S\_avg}^{\lceil\sqrt{n}\rceil}\left(D'^c_{\boldsymbol{F}_{\boldsymbol{i}}}(y^n)|\boldsymbol{c}\right)\\
     &\le 2^{-\lceil\sqrt{n}\rceil\delta}=\circ(n).
\end{align*}
Thus, we can achieve an arbitrary small type I error probability with sufficient large code length.

Without the loss of generality, we examine the type II error probability $P^1_{e,2}$ for the identity set $\mathcal{N}_1$. We first introduce the definition of the following sets for all $\boldsymbol{i},\boldsymbol{i}'\in\boldsymbol{\mathcal{N}}$.
\begin{align*}
    F_{i_1}\cap F_{i'_1}&:= \left\{y\in\mathcal{D}^*|F_{1,i_1}(y^n)=F_{1,i'_1}(y^n)\right\},\\
    F_{i/i_1}\cap F_{i'/i'_1}&:= \left\{y\in\mathbb{R}|\boldsymbol{F}_{\boldsymbol{i/i_1}}(y^n)=\boldsymbol{F}_{\boldsymbol{i'/i'_1}}(y^n)\right\},\\
    F_{i_1}- F_{i'_1}&:= \left\{y\in\mathcal{D}^*|F_{1,i_1}(y^n)\ne F_{1,i'_1}(y^n)\right\},\\
    F_{i/i_1}- F_{i'/i'_1}&:= \left\{y\in\mathcal{D}^*|\boldsymbol{F}_{\boldsymbol{i/i_1}}(y^n)\ne\boldsymbol{F}_{\boldsymbol{i'/i'_1}}(y^n)\right\},
\end{align*}
where $\boldsymbol{i/i_1}=(i_2,\cdots,i_K)$ and $\boldsymbol{i'/i'_1}=(i'_2,\cdots,i'_K)$.

Then, for all $\boldsymbol{i},\boldsymbol{\tilde{i}}\in\boldsymbol{\mathcal{N}}$, $\boldsymbol{i}\ne\boldsymbol{\tilde{i}}$, the type II error probability $P_{e2}^1(\boldsymbol{i},\boldsymbol{\tilde{i}})$ at Sender 1 can be upper-bounded by:
\begin{align*}
    &W_{S\_avg}^{m}(\mathcal{D}_{\boldsymbol{\tilde{i}_1}}|\boldsymbol{f}_{\boldsymbol{i}})\\
    &\le \sum_{y^n\in \{F_{i_1}\cap F_{\tilde{i}_1}\}\cap \{F_{i/i_1}\cap F_{i'/i'_1}\}} W_{S\_avg}^n(y^n|\boldsymbol{x}^{*n})\\
    &\quad +\sum_{y^n\in \{F_{i_1}-F_{\tilde{i}_1}\}\cup \{F_{i/i_1}-F_{i'/i'_1}\}} W_{S\_avg}^n(y^n|\boldsymbol{x}^{*n})\cdot 2^{-\lceil\sqrt{n}\rceil\delta}\\
    &\le\frac{|F_{i_1}\cap F_{\tilde{i}_1}|}{|\mathcal{D}^*|}+\circ(n).
\end{align*}
Define an auxiliary random variable $\Psi_{y^n}$ for all $y^n\in \mathcal{D}^*$, where
\begin{align*}
    \Psi_{y^n}(F_{\tilde{i}})=\left\{
        \begin{array}{cc}
             1,&y^n\in F_{i_1}\cap F_{\tilde{i}_1}  \\
             0,&y^n\in F_{i_1}-F_{\tilde{i}_1}
        \end{array},
    \right.
\end{align*}
with probability $Pr[\Psi_{y^n}(F_{\tilde{i}})=1]=\frac{1}{M_1}$.
\begin{lemma}\cite{ahlswede1989identificationfeedback}
    For $\lambda \in(0,1)$, and $E[\Psi_{y^n}]=\frac{1}{M_1}<\lambda$,
    \begin{align*}
        Pr\left[\frac{1}{|\mathcal{D}^*|}\sum_{y^n\in \mathcal{D}^*}\Psi_{y^n}(F_2)>\lambda\right]
        <2^{-|\mathcal{D}^*|\cdot (\lambda\log(M_1)-1)}.
    \end{align*}
\end{lemma}
For all pairs $(i_1,\tilde{i}_1)\in\mathcal{N}_1^2$, $i_1\ne\tilde{i}_1$, we need to upper-bound $\frac{|F_{i_1}\cap F_{\tilde{i}_1}|}{|\mathcal{D}^*|}$ by $\lambda$. Thus, the following probability
\begin{align*}
    &Pr\left[ \bigcap_{\tilde{i}\in\mathcal{N}_1,\tilde{i}\ne i}\left\{\frac{1}{|\mathcal{D}^*|}\sum_{y^n\in \mathcal{D}^*}\Psi_{y^n}(F_{\tilde{i}})\le\lambda\right\}\right]\\
    &\quad \ge 1- (N_1-1)\cdot 2^{-2^{nH(W_{S\_avg}(\cdot|\boldsymbol{x}^*)}\cdot(\lambda\log{M_1}-1)}
\end{align*}
should be greater than $0$. Therefore, the maximum value of $N_1$ we can choose is:
\begin{align*}
    N_1=2^{2^{nH(W_{S\_avg}(\cdot|\boldsymbol{x}^*))}\cdot(\lambda\log{M_1}-1)}. 
\end{align*}
This implies that as $m$ goes to infinity, the rate at Sender 1 can achieve:
\begin{align*}
    \lim_{m\to\infty}R_1(\boldsymbol{D})
    &=\lim_{m\to\infty}\frac{\log\log{N_1}}{m}=H(W_{S\_avg}(\cdot|\boldsymbol{x}^*)).
\end{align*}
Similarly, if the rate at any Sender $k\in\mathcal{K}$ takes the value
\begin{align*}
    \lim_{m\to\infty}R_k(\boldsymbol{D})= H(W_{S\_avg}(\cdot|\boldsymbol{x}^*)),
\end{align*}
then the type II error probability at Sender $k$ can also be upper-bounded by $\lambda$.
Regarding distortion, our coding scheme guarantees:
\begin{align*}
    \bar{d}_k^m
    &=\frac{1}{N_k}\sum_{i_k=1}^{N_k}\frac{1}{m}\left(n\cdot d_k^*(x_k^*)+\lceil\sqrt{n}\rceil\cdot\bar{d}(\boldsymbol{c}_k\left(F_i(y^n\right)))\right)\le D_k,
\end{align*}
This completes the proof of Theorem \ref{thm.JdIDAS}.
\subsection{Sketch of Proof of Theorem \ref{thm.JrIDAS}}
Similarly, consider a randomized IDF code with code length $m=n+\lceil\sqrt{n}\rceil$. The first $n$ bits are generated according to the same probability distribution $\boldsymbol{P}^*$, where
\begin{align*}
    \boldsymbol{P}^*=\underset{{\boldsymbol{P}\in\mathcal{P}(\boldsymbol{\mathcal{X}})}}{\arg\max}H(\boldsymbol{P}W_{S\_avg}).
\end{align*}
We convert it to an uniformly distributed random experiment $\left(\mathcal{D}^*,\boldsymbol{P}^*W_{S\_avg}\right)$, where $\mathcal{D}^*=\mathcal{T}_\epsilon^n(\boldsymbol{P}^*W_{S\_avg})$.
The consequent encoding steps are similar to the deterministic case.\\
In conclusion, our randomized IDF code can be represented as follows:
\begin{align*}
    Q_F(\boldsymbol{f}_{k}|i_k)&={\boldsymbol{P}_k^*}^{n}\times \mathbbm{1}\left(\boldsymbol{f}_{k,n+1}^m=c_k\right),
    \\P_k^*&\in\mathcal{P}_k^{D_k},\bar{d}_k(\boldsymbol{c}_k)\le D_k, \quad \forall k\in\mathcal{K},\\
    \mathcal{D}_{\boldsymbol{i}} &= \bigcup_{y^n\in \mathcal{D}^*}y^n\times \mathcal{D}'_{\boldsymbol{F}_{\boldsymbol{i}}(y^n)},\quad \forall \boldsymbol{i}\in\boldsymbol{\mathcal{N}}.
\end{align*}
Similarly, we can achieve an arbitrary small type I error probability with sufficient large code length. With
\begin{align*}
    \lim_{m\to\infty}R_{k}(\boldsymbol{D})=\max_{\boldsymbol{P}\in\boldsymbol{\mathcal{P}}^{\boldsymbol{D}}}H(\boldsymbol{P}W_{S\_avg})
\end{align*}
we can upper-bound the probabilities of type II error at any Sender $K$ by $\lambda$.\\
The distortion constraints are satisfied as follows:
\begin{align*}
    \bar{d}_k^m
    &=\frac{1}{N_k}\sum_{i_k=1}^{N_k}\frac{1}{m}\left(n\cdot d_k^*(P_k^*)+\lceil\sqrt{n}\rceil\cdot\bar{d}_k(\boldsymbol{c}_k(\boldsymbol{F}_{\boldsymbol{i}}(y^n)))\right)\le D_k.
\end{align*}
This completes the proof of Theorem \ref{thm.JrIDAS}.

%========================== Examples ======================
\section{Example}
\label{sec: examples}
%For deterministic ID, the optimization problem can be written as
%\begin{align*}
%    \text{maximize} \quad\quad  &J^d=H(W_{S\_avg}(\cdot|x_1,x_2))\\
%    \text{subject to} \quad \quad &d_1^*(x_1)\le D_1\\
%    &d_2^*(x_2)\le D_2.
%\end{align*}
%For randomized ID, the optimization problem is stated as follows:
%\begin{align*}
%    \text{maximize} \quad \quad &J^r(P_1,P_2)=H(P_1P_2W_{S\_avg})\\
%    \text{subject to} \quad \quad &d_1^*(P_1)\le D_1\\
%    &d_2^*(P_2)\le D_2\\
%    &\sum_{x_1\in\mathcal{X}_1}P_1(x_1)=1\\
%    &\sum_{x_2\in\mathcal{X}_2}P_2(x_2)=1.
%\end{align*}
Consider a binary adder 2-SD-MAC with distributed multiplicative Bernoulli states, where $ Y=X_1\cdot S_1\oplus X_2\cdot S_2$. We assume $\mathcal{X}_1=\mathcal{X}_2=\mathcal{S}_1=\mathcal{S}_2=\mathcal{Y}=\{0,1\}$, and $S_1,S_2\sim Ber(p_S)$ are i.i.d. for all $t=1,\cdots,n$. Hamming distance is used as the distortion function, i.e., $d(S_k,\hat{S}_k)=S_k\oplus \hat{S}_k$.
The estimates are given by
\begin{align*}
    \hat{s}_k=h_k^*(x_k,y)=\underset{s'_k\in\{0,1\}}{\arg\max}P_{S_k|X_kY}(s'_k|x_k,y).
\end{align*}
For deterministic ID, the minimal distortions of $x_k$ are given by
\begin{align*}
    d_k^*(x_k)&=\sum_{y}P_{Y|X_k}(y|x_k)\min_{s''_k\in\{0,1\}}P_{S_k|X_kY}(s''_k|x_k,y).
\end{align*}
The extreme points are listed in TABLE \ref{tab.extreme points}.
\begin{table}[ht]
    \centering
    \renewcommand{\arraystretch}{1.5} 
    \setlength{\tabcolsep}{6pt}
    \scalebox{0.75}{
    \begin{tabular}{|c c|c|c c|c|}
        \hline
        $x_1$ & $x_2$ & $Y$ & $d_1^{*}(x_1)$ & $d_2^{*}(x_2)$ & $R_1,R_2\ge$ \\
        \hline
        $0$ & $1$ & $S_2$ & $\min\{p_S,1-p_S\}$ & $0=D_{2,min}$ & $H_2(p_S)$ \\
        \hline
        $1$ & $0$ & $S_1$ & $0=D_{1,min}$ & $\min\{p_S,1-p_S\}$ & $H_2(p_S)$ \\
        \hline
        $1$ & $1$ & $S_1\oplus S_2$ & $p_S=D_{1,max}$ & $p_S=D_{2,max}$ & $H_2\left(2p_S(1-p_S)\right)=R_{max}$ \\
        \hline
    \end{tabular}
    }
    \caption{Extreme points of lower-bound of deterministic ID capacity-distortion region}
    \label{tab.extreme points}
    \vspace{-1em}
\end{table}

\begin{figure}
\centering
\subfloat[Lower bound of $\mathcal{C}^r_{ID}(\boldsymbol{D})$with $p_S=0.2$]{\input{figures/example}
\label{fig:image1}}\\
\subfloat[Lower bound of $C^r_k(\boldsymbol{D})$ with assumption $D_1=D_2$]{\includegraphics[width=0.36\textwidth]{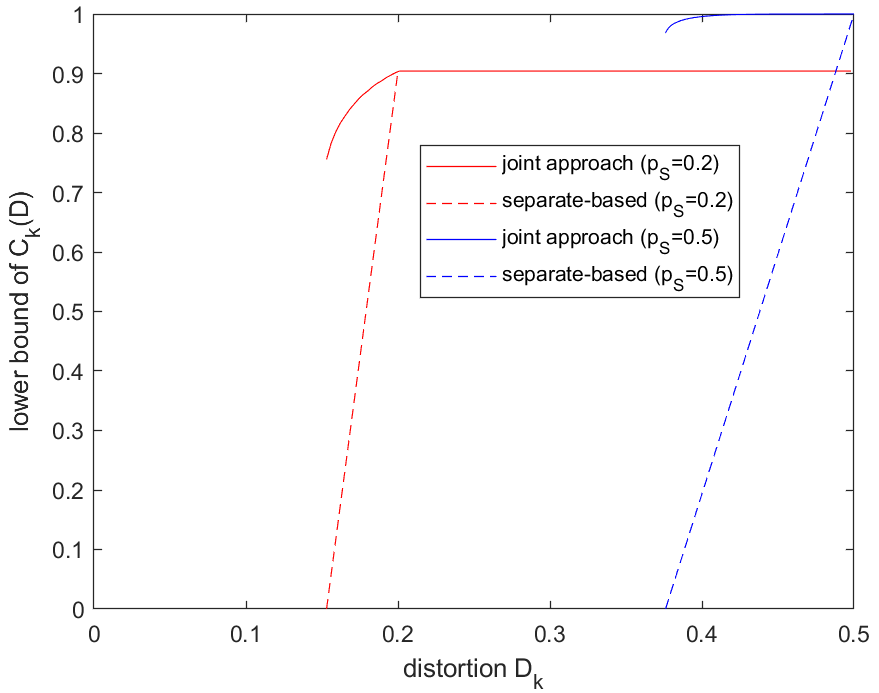}}
\vspace{-0.5em}
\caption{Joint approach vs. separation-based approach}
\vspace{-1.5em}
\label{fig:tradeoff}
\end{figure}

\noindent For randomized ID, we assume $X_1\sim Ber(p_1)$ and $X_2\sim Ber(p_2)$. The minimal distortions of $P_k$ are given by
\begin{align*}
    d_k^*(p_k)=(1-p_k)d^*(0)+p_kd^*(1).
\end{align*}
Our lower bounds of the randomized ID capacity-distortion region are illustrated in Fig. \ref{fig:tradeoff}. It is evident that our joint approach results in a significant enhancement compared to the separation-based method achieved by splitting the resources either estimating the SI or communication, especially when $p_S\approx\frac{1}{2}$. Note that not every pair $(D_1,D_2)$ is achievable.
%===================== conclusions ========================

\section{Conclusions}
\label{sec: conclusion}
In our work, we studied the problem of joint ID and channel state sensing over a 2-SD-MAC where two i.i.d. channel states are generated jointly. Each sender is equipped with an estimator tasked with sensing the respective channel state based on input and feedback. Our contribution lies in establishing lower bounds for both deterministic and randomized ID capacity-distortion regions in this setup. Future research directions could focus on establishing the converse proof of the JIDAS problem. Additionally, extending this problem to Gaussian channels would also be interesting.

\section*{Acknowledgments}
H. Boche, C. Deppe, W. Labidi, and Y. Zhao acknowledge the financial support by the Federal Ministry of Education and Research
of Germany (BMBF) in the program of “Souverän. Digital. Vernetzt.”. Joint project 6G-life, project identification number: 16KISK002.
Eduard A. Jorswieck is supported by
the Federal Ministry of Education and Research (BMBF, Germany) through
the Program of “Souveran. Digital. Vernetzt.” Joint Project 6G-Research and
Innovation Cluster (6G-RIC) under Grant 16KISK031
H. Boche and W. Labidi were further supported in part by the BMBF within the national initiative on Post Shannon Communication (NewCom) under Grant 16KIS1003K. C.\ Deppe was further supported in part by the BMBF within NewCom under Grant 16KIS1005. C. Deppe, W. Labidi and Y. Zhao were also supported by the DFG within the project DE1915/2-1. 
\clearpage
\newpage
\bibliographystyle{IEEEtran}
\bibliography{definitions,references}

\begin{thebibliography}{10}
\providecommand{\url}[1]{#1}
\csname url@samestyle\endcsname
\providecommand{\newblock}{\relax}
\providecommand{\bibinfo}[2]{#2}
\providecommand{\BIBentrySTDinterwordspacing}{\spaceskip=0pt\relax}
\providecommand{\BIBentryALTinterwordstretchfactor}{4}
\providecommand{\BIBentryALTinterwordspacing}{\spaceskip=\fontdimen2\font plus
\BIBentryALTinterwordstretchfactor\fontdimen3\font minus
  \fontdimen4\font\relax}
\providecommand{\BIBforeignlanguage}[2]{{%
\expandafter\ifx\csname l@#1\endcsname\relax
\typeout{** WARNING: IEEEtran.bst: No hyphenation pattern has been}%
\typeout{** loaded for the language `#1'. Using the pattern for}%
\typeout{** the default language instead.}%
\else
\language=\csname l@#1\endcsname
\fi
#2}}
\providecommand{\BIBdecl}{\relax}
\BIBdecl

\bibitem{shannon1948mathematical}
C.~E. Shannon, ``A mathematical theory of communication,'' \emph{The Bell
  system technical journal}, vol.~27, no.~3, pp. 379--423, 1948.

\bibitem{boche2018secure}
H.~Boche and C.~Deppe, ``Secure identification for wiretap channels;
  robustness, super-additivity and continuity,'' \emph{IEEE Transactions on
  Information Forensics and Security}, vol.~13, no.~7, pp. 1641--1655, 2018.

\bibitem{moulin2001role}
P.~Moulin, ``The role of information theory in watermarking and its application
  to image watermarking,'' \emph{Signal Processing}, vol.~81, no.~6, pp.
  1121--1139, 2001.

\bibitem{ahlswede2006watermarking}
R.~Ahlswede and N.~Cai, ``Watermarking identification codes with related topics
  on common randomness,'' \emph{General Theory of Information Transfer and
  Combinatorics}, pp. 107--153, 2006.

\bibitem{steinberg2001identification}
Y.~Steinberg and N.~Merhav, ``Identification in the presence of side
  information with application to watermarking,'' \emph{IEEE Transactions on
  Information Theory}, vol.~47, no.~4, pp. 1410--1422, 2001.

\bibitem{lu2017industry}
Y.~Lu, ``Industry 4.0: A survey on technologies, applications and open research
  issues,'' \emph{Journal of industrial information integration}, vol.~6, pp.
  1--10, 2017.

\bibitem{fettweis20226g}
G.~P. Fettweis and H.~Boche, ``On {6G} and trustworthiness,''
  \emph{Communications of the ACM}, vol.~65, no.~4, pp. 48--49, 2022.

\bibitem{cabrera20216g}
J.~A. Cabrera, H.~Boche, C.~Deppe, R.~F. Schaefer, C.~Scheunert, and F.~H.
  Fitzek, ``{6G} and the post-{Shannon} theory,'' \emph{Shaping Future 6G
  Networks: Needs, Impacts, and Technologies}, pp. 271--294, 2021.

\bibitem{ahlswede1989identification}
R.~Ahlswede and G.~Dueck, ``Identification via channels,'' \emph{IEEE
  Transactions on Information Theory}, vol.~35, no.~1, pp. 15--29, 1989.

\bibitem{ja1985identification}
J.~J. Ja, ``Identification is easier than decoding,'' in \emph{26th Annual
  Symposium on Foundations of Computer Science (sfcs 1985)}.\hskip 1em plus
  0.5em minus 0.4em\relax IEEE, 1985, pp. 43--50.

\bibitem{ahlswede1995new}
R.~Ahlswede and Z.~Zhang, ``New directions in the theory of identification via
  channels,'' \emph{IEEE transactions on information theory}, vol.~41, no.~4,
  pp. 1040--1050, 1995.

\bibitem{labidi2020secure}
W.~Labidi, C.~Deppe, and H.~Boche, ``Secure identification for {Gaussian}
  channels,'' in \emph{ICASSP 2020-2020 IEEE International Conference on
  Acoustics, Speech and Signal Processing (ICASSP)}.\hskip 1em plus 0.5em minus
  0.4em\relax IEEE, 2020, pp. 2872--2876.

\bibitem{boche2019secure}
H.~Boche, C.~Deppe, and A.~Winter, ``Secure and robust identification via
  classical-quantum channels,'' \emph{IEEE Transactions on Information Theory},
  vol.~65, no.~10, pp. 6734--6749, 2019.

\bibitem{pereg2022identification}
U.~Pereg, J.~Rosenberger, and C.~Deppe, ``Identification over quantum broadcast
  channels,'' in \emph{2022 IEEE International Symposium on Information Theory
  (ISIT)}.\hskip 1em plus 0.5em minus 0.4em\relax IEEE, 2022, pp. 258--263.

\bibitem{ezzine2024common}
R.~Ezzine, M.~Wiese, C.~Deppe, and H.~Boche, ``Common randomness generation
  from finite compound sources,'' \emph{arXiv preprint arXiv:2401.14323}, 2024.

\bibitem{ezzine2021common}
------, ``Common randomness generation over slow fading channels,'' in
  \emph{2021 IEEE International Symposium on Information Theory (ISIT)}.\hskip
  1em plus 0.5em minus 0.4em\relax IEEE, 2021, pp. 1925--1930.

\bibitem{labidi2022common}
W.~Labidi, R.~Ezzine, C.~Deppe, and H.~Boche, ``Common randomness generation
  from gaussian sources,'' \emph{arXiv preprint arXiv:2201.11078}, 2022.

\bibitem{ahlswede1989identificationfeedback}
R.~Ahlswede and G.~Dueck, ``Identification in the presence of feedback-a
  discovery of new capacity formulas,'' \emph{IEEE Transactions on Information
  Theory}, vol.~35, no.~1, pp. 30--36, 1989.

\bibitem{wolfowitz2012coding}
J.~Wolfowitz, \emph{Coding theorems of information theory}.\hskip 1em plus
  0.5em minus 0.4em\relax Springer Science \& Business Media, 2012, vol.~31.

\bibitem{Schwenteck2023}
P.~Schwenteck, G.~T. Nguyen, H.~Boche, W.~Kellerer, and F.~H.~P. Fitzek, ``6g
  perspective of mobile network operators, manufacturers, and verticals,''
  \emph{{IEEE} Network Letters}, vol.~5, no.~3, pp. 169--172, 2023.

\bibitem{bourdoux20206g}
A.~Bourdoux, A.~N. Barreto, B.~van Liempd, C.~de~Lima, D.~Dardari, D.~Belot,
  E.-S. Lohan, G.~Seco-Granados, H.~Sarieddeen, H.~Wymeersch \emph{et~al.},
  ``{6G} white paper on localization and sensing,'' \emph{arXiv preprint
  arXiv:2006.01779}, 2020.

\bibitem{Fettweis2021}
G.~P. Fettweis and H.~Boche, ``6g: The personal tactile internet—and open
  questions for information theory,'' \emph{{IEEE} {BITS} the Information
  Theory Magazine}, vol.~1, no.~1, pp. 71--82, 2021.

\bibitem{zheng2019radar}
L.~Zheng, M.~Lops, Y.~C. Eldar, and X.~Wang, ``Radar and communication
  coexistence: An overview: A review of recent methods,'' \emph{IEEE Signal
  Processing Magazine}, vol.~36, no.~5, pp. 85--99, 2019.

\bibitem{liu2020joint}
F.~Liu, C.~Masouros, A.~P. Petropulu, H.~Griffiths, and L.~Hanzo, ``Joint radar
  and communication design: Applications, state-of-the-art, and the road
  ahead,'' \emph{IEEE Transactions on Communications}, vol.~68, no.~6, pp.
  3834--3862, 2020.

\bibitem{sturm2011waveform}
C.~Sturm and W.~Wiesbeck, ``Waveform design and signal processing aspects for
  fusion of wireless communications and radar sensing,'' \emph{Proceedings of
  the IEEE}, vol.~99, no.~7, pp. 1236--1259, 2011.

\bibitem{bliss2014cooperative}
D.~W. Bliss, ``Cooperative radar and communications signaling: The estimation
  and information theory odd couple,'' in \emph{2014 IEEE Radar
  Conference}.\hskip 1em plus 0.5em minus 0.4em\relax IEEE, 2014, pp.
  0050--0055.

\bibitem{bica2015opportunistic}
M.~Bica, K.-W. Huang, U.~Mitra, and V.~Koivunen, ``Opportunistic radar waveform
  design in joint radar and cellular communication systems,'' in \emph{2015
  IEEE Global Communications Conference (GLOBECOM)}.\hskip 1em plus 0.5em minus
  0.4em\relax IEEE, 2015, pp. 1--7.

\bibitem{huang2015radar}
K.-W. Huang, M.~Bic{\u{a}}, U.~Mitra, and V.~Koivunen, ``Radar waveform design
  in spectrum sharing environment: Coexistence and cognition,'' in \emph{2015
  IEEE Radar Conference (RadarCon)}.\hskip 1em plus 0.5em minus 0.4em\relax
  IEEE, 2015, pp. 1698--1703.

\bibitem{kobayashi2018joint}
M.~Kobayashi, G.~Caire, and G.~Kramer, ``Joint state sensing and communication:
  Optimal tradeoff for a memoryless case,'' in \emph{2018 IEEE International
  Symposium on Information Theory (ISIT)}.\hskip 1em plus 0.5em minus
  0.4em\relax IEEE, 2018, pp. 111--115.

\bibitem{9834554}
M.~Ahmadipour, M.~Wigger, and M.~Kobayashi, ``Coding for sensing: An improved
  scheme for integrated sensing and communication over macs,'' in \emph{2022
  IEEE International Symposium on Information Theory (ISIT)}, 2022, pp.
  3025--3030.

\bibitem{10153971}
M.~Ahmadipour and M.~Wigger, ``An information-theoretic approach to
  collaborative integrated sensing and communication for two-transmitter
  systems,'' \emph{IEEE Journal on Selected Areas in Information Theory},
  vol.~4, pp. 112--127, 2023.

\bibitem{labidi2023joint}
W.~Labidi, C.~Deppe, and H.~Boche, ``Joint identification and sensing for
  discrete memoryless channels,'' in \emph{2023 IEEE International Symposium on
  Information Theory (ISIT)}.\hskip 1em plus 0.5em minus 0.4em\relax IEEE,
  2023, pp. 442--447.

\bibitem{liao1972multiple}
H.~H.-J. Liao, ``Multiple access channels,'' Ph.D. dissertation, University of
  Hawaii Honolulu, HI, USA, 1972.

\bibitem{el2011network}
A.~El~Gamal and Y.-H. Kim, \emph{Network information theory}.\hskip 1em plus
  0.5em minus 0.4em\relax Cambridge university press, 2011.

\bibitem{rosenberger2023deterministic}
J.~Rosenberger, A.~Ibrahim, C.~Deppe, and R.~Ferrara, ``Deterministic
  identification over multiple-access channels,'' in \emph{2023 IEEE
  International Symposium on Information Theory (ISIT)}, 2023.

\bibitem{ahlswede2008general}
R.~Ahlswede, ``General theory of information transfer: Updated,''
  \emph{Discrete Applied Mathematics}, vol. 156, no.~9, pp. 1348--1388, 2008.

\bibitem{diadamo2019simultaneous}
S.~Diadamo and H.~Boche, ``The simultaneous identification capacity of the
  classical--quantum multiple access channel with stochastic encoders for
  transmission,'' \emph{arXiv preprint arXiv:1903.03395}, 2019.

\bibitem{ahlswede1971multi}
R.~Ahlswede, ``Multi-way communication channels,'' in \emph{Proc. 2nd. Int.
  Symp. Information Theory (Tsahkadsor, Armenian SSR), 1971}.\hskip 1em plus
  0.5em minus 0.4em\relax Publishing House of the Hungarian Academy of
  Sciences, 1971, pp. 23--52.

\end{thebibliography}
\IEEEtriggeratref{4}
%%
%% which triggers a \newpage (i.e., new column) just before the given
%% reference number. Note that you need to adapt this if you modify
%% the paper.  The "triggered" command can be changed if desired:
%%
%\IEEEtriggercmd{\enlargethispage{-20cm}}
%%
%%%%%%
%\section*{References}
%\bibliographystyle{elsarticle-num}

%%%%%%
%% References:
%% We recommend the usage of BibTeX:
%%
%\bibliographystyle{IEEEtran}
%\bibliography{definitions,bibliofile}
%%
%% where we here have assume the existence of the files
%% definitions.bib and bibliofile.bib.
%% BibTeX documentation can be obtained at:
%% http://www.ctan.org/tex-archive/biblio/bibtex/contrib/doc/
%%%%%%

%% Or you use manual references (pay attention to consistency and the
%% formatting style!):

\end{document}